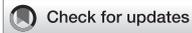

# Investigation on different materials after pulsed high field conditioning and low-energy H⁻ irradiation

C. Serafim[1,2]\*, R. Peacock[1], S. Calatroni[1], F. Djurabekova[2], A. T. Perez Fontenla[1], W. Wuensch[1], S. Sgobba[1], A. Grudiev[1], A. Lombardi[1], E. Sargsyan[1], S. Ramberger[1] and G. Bellodi[1]

[1]CERN Organisation Européenne pour la Recherche Nucléaire, Geneva, Switzerland, [2]Department of Physics, University of Helsinki, Helsinki, Finland

During operation, the radio-frequency quadrupole (RFQ) of the LINAC4 at CERN is exposed to high electric fields, which can lead to vacuum breakdown. It is also subject to beam loss, which can cause surface modification, including blistering, which can result in reduced electric field holding and an increased breakdown rate. First, experiments to study the high-voltage conditioning process and electrical breakdown statistics have been conducted using pulsed high-voltage DC systems in order to identify materials with high electric field handling capability and robustness to low-energy irradiation. In this paper, we discuss the results obtained for the different materials tested. To complement these, an investigation of their metallurgical properties using advanced microscopic techniques was done to observe and characterize the different materials and to compare results before and after irradiation and breakdown testing.

## 1 Introduction

The first stage of acceleration in the Large Hadron Collider (LHC) at CERN is a linear accelerator known as LINAC4 [1]. H⁻ ions are generated by the radio-frequency (RF) source and pass through the low-energy beam transport (LEBT) to the radio-frequency quadrupole (RFQ). An endoscopy of the LINAC4 RFQ showed damage of the copper surfaces and signs of breakdown, in particular at its entrance where high beam losses can be expected [2]. To better understand this problem, tests were conducted to gain more knowledge of the effects of low-energy irradiation on field holding capabilities, on possible surface blistering, and its consequences on the breakdown phenomena. A vacuum breakdown is a discharge across a gap in a circuit, typically occurring when the voltage between two conductive electrodes becomes high enough to ionize the surroundings and allow the flow of current. In the case of our system, a breakdown can be triggered by applying a potential difference between the two electrodes separated by a given gap, under vacuum. When applying a high voltage potential, if the electric field on the electrode surface is high enough, the emissions of neutrals and their sequential ionization will form a conductive plasma within our gap. This plasma will serve as a bridge between the two electrode surfaces, and the emission of electrons from the cathode surface will result in an exchange of charge between the two electrodes [3]. More detailed information about the process behind the formation of a vacuum breakdown has been reported in multiple research works [3–6]. The vacuum breakdowns from this process are visually observed as craters on the material surface with an order of magnitude of tens of micrometers. Their size of the craters is directly dependent on the magnitude of the electric field applied.

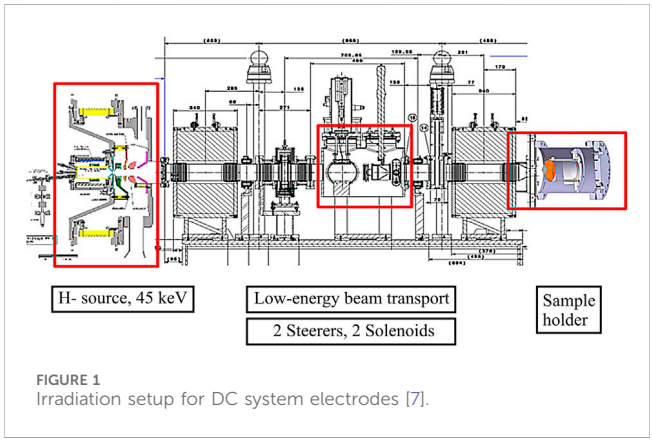

FIGURE 1
Irradiation setup for DC system electrodes [7].

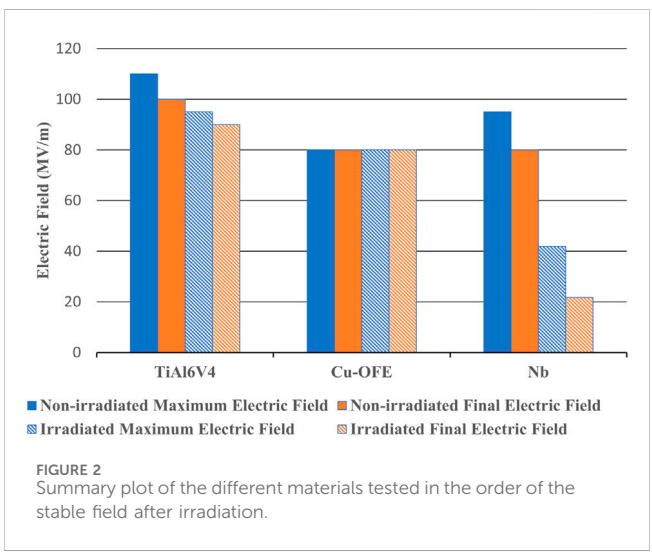

FIGURE 2
Summary plot of the different materials tested in the order of the stable field after irradiation.

The study of the breakdown phenomenon in linear colliders is highly important, as an increase in the accelerating gradients, and consequently, the increase in the electric fields on the base–material surfaces will increase the probability of the occurrence of a breakdown. For a better understanding whether beam irradiation could have a consequent effect in triggering breakdowns, tests were conducted on pairs of anode and cathode electrodes of different materials that were manufactured for the purpose of the studies. Their breakdown handling capabilities were tested in a dedicated pre-existing pulsed high-voltage set-up [4], the large electrode system (LES). Detailed information about the test conditions can be found in Section 2. For comparison, a pair of non-irradiated electrodes of each selected material was also tested in this system.

For the study of the effects of irradiation, and of the material properties at its origin, a test stand that replicates the low energy beam transport in the LINAC4 tunnel at the CERN was used, as seen in Figure 1. In this test stand, a specific hardware was developed to use a cathode as the target for irradiation and allowing for a rapid turnover for various irradiation runs.

Irradiating a material with H⁻ ions can lead to various effects on the surface material. The outcomes depend on factors such as the type of the material, the ion dose, and the energy level of the ions. The irradiation may cause changes in the surface morphology and composition of the metal, as well as induce structural and chemical changes in the metal lattice. Ion implantation has been reported to enhance properties such as hardness, wear resistance, or corrosion resistance [8]. Depending on the metal type, irradiation with hydrogen ions can lead to hydrogen embrittlement. Hydrogen atoms can diffuse into the metal lattice, affecting its mechanical properties and making it more susceptible to cracking and fracture [9, 10]. Depending on the diffusibility and solubility factors of each material, H⁻ ions will have different rates of diffusion into the material lattice, and consequently, some materials can have surface modification from the irradiation, with blistering appearance, increasing the roughness of the surface, as we have seen in copper. Surface modification of materials due to the blistering phenomenon caused by hydrogen retention is a known problem [11]. Blistering is dependent on many factors, such as the radiation dose (minimum dose required for the appearance of blisters), the ion energy, the temperature of the target, the angle of incidence [7], and the metallurgical properties of the material. In the case of the RFQ, the energy of the incident H⁻ beam at its entrance, where most beam losses are localized, is 45 keV. This corresponds to a range of approximately 250 nm in copper, with the H possibly accumulating under the surface and leading to blistering. All electrodes underwent systematic metallurgical analyses, before and after irradiation, and after LES testing, as detailed in this paper.

## 2 Methodology

A list of candidate materials was carefully drafted for the manufacturing of a future RFQ with better performance. In this paper, we will focus on the results of three different materials: oxygen free electronic-grade copper (Cu OFE, UNS C10100) as the reference employed for the current RFQ fabrication, high purity (99.9%, RRR300) niobium (Nb), and a premium-grade titanium alloy (Ti6Al4V, grade F-5, 6 wt% aluminum—4 wt% vanadium, UNS R56400), forged in an α–β range with a final α-microstructure.

The materials were selected based on their usability for meter-long high gradient RF cavities and their potential resistance to blistering and breakdown phenomena. Nb and Ti grade 5 were chosen in particular for their high hydrogen diffusivity, which should prevent accumulation of hydrogen and thus blistering. Ti grade 5 should also provide comparatively easier machinability. The machining of the electrodes was performed with high precision tooling. After machining, the electrodes were submitted to metrology testing to assure that all the requirements are within the stipulated tolerances. After this, the electrodes have been submitted to a cleaning process. For the case of Cu, a thermal treatment equivalent to the brazing cycle employed for RFQ fabrication was applied, with a heat up ramp rate of 100°/h up to 795°C, dwell time of 6 h at 795°C and 100°/h up to 835°C, and dwell

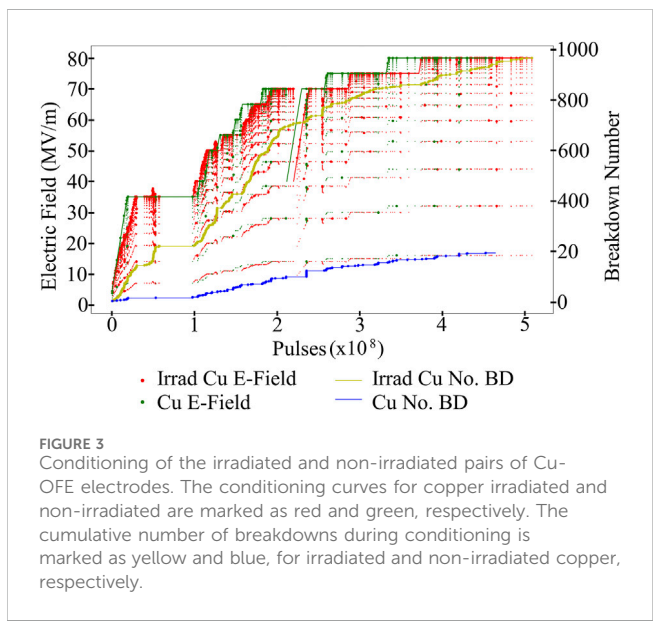

FIGURE 3
Conditioning of the irradiated and non-irradiated pairs of Cu-OFE electrodes. The conditioning curves for copper irradiated and non-irradiated are marked as red and green, respectively. The cumulative number of breakdowns during conditioning is marked as yellow and blue, for irradiated and non-irradiated copper, respectively.

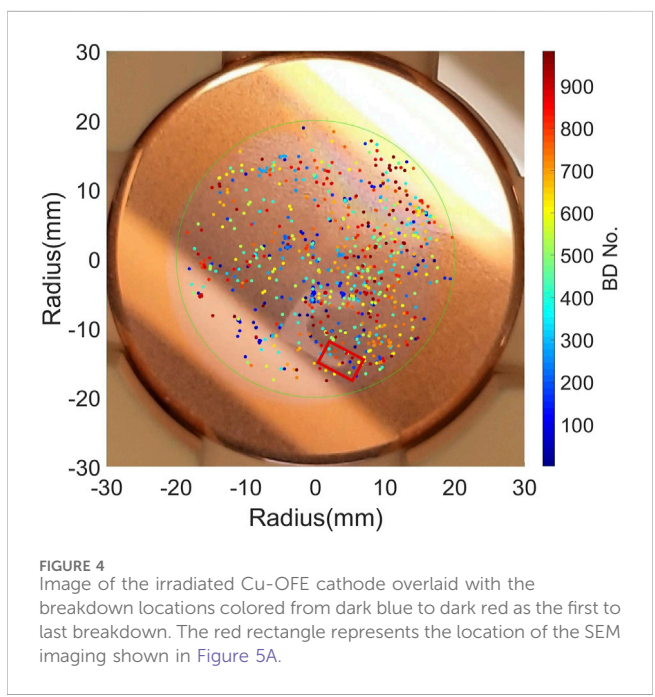

FIGURE 4
Image of the irradiated Cu-OFE cathode overlaid with the breakdown locations colored from dark blue to dark red as the first to last breakdown. The red rectangle represents the location of the SEM imaging shown in Figure 5A.

time of 45 min at 835°C. The natural cooling rate was used in the furnace.

Irradiation was performed at CERN using a low-energy H⁻ beam at 45 keV. Each irradiation had a duration of approximately 52–54 h, with a beam pulse duration of 600 μs, a repetition rate of 0.83 Hz, and a peak pulse current of 20 mA, which resulted in a deposition of $1.2 \times 10^{19}$ H⁻p/cm² on the target, corresponding to approximately 10 days of beam losses during RFQ operation [12].

The high-voltage testing was performed on the electrodes exposed to irradiation and on electrodes that were not exposed to irradiation. This allowed comparing whether the irradiation had a significant impact on the performance of each material in reaching a maximum and stable field.

The pulsed DC LES consists of two high-precision machined electrodes placed 60 μm apart in vacuum on the order of $1 \times 10^{-8}$ mbar. A Marx generator was used to apply pulses of voltage up to 10 kV, pulse lengths between 1 μs and 1 ms, and repetition rate up to 6 kHz [3]. This system is used for the study of conditioning and vacuum breakdown phenomena. The instrumentation has the ability to detect breakdowns using voltage and current signals, measure pressure spikes, and emitted light used to determine the location of each breakdown during operation [13].

To replicate conditions occurring in the RFQ as closely as possible while keeping testing time to a minimum, a pulse length of 100 μs and a repetition rate of 200 Hz were used. Once conditioned, pulse length dependence tests were also done up to the pulse length of the RFQ of up to 900 μs; these did not show a clear dependence of pulse length on the breakdown rate for pulse lengths over 100 μs [14]. To replicate the electric field of 34 MV/m of the RFQ, the LES was first conditioned to 35 MV/m before increasing the field in steps to find the limit [2]. The gap dependence effect [15] is a possible explanation for the higher electric fields achieved in the LES over the RFQ. The summary plot in Figure 2 shows the materials tested, the maximum field reached, and the final stable field.

In order to inspect the state of the electrodes between each test, microscopy equipment was used. The samples were analyzed by using a digital microscope from KEYENCE and a scanning electron microscope (SEM) (Sigma) from Zeiss, equipped with secondary electron (SE), InLens (Inlens), Everhart–Thornley SE (SE2), backscattered electron (AsB), and electron backscattered diffraction (EBSD) detectors. The results from the observation help in understanding the development of blistering, if any, created on the surface from the irradiation and a possible correlation with the location of the breakdowns from the LES.

## 3 Results

### 3.1 Oxygen-free electronic copper (Cu-OFE)

Cu-OFE achieved the second highest stable electric field of the materials tested after irradiation, achieving 80 MV/m. The non-irradiated pair achieved the same stable electric field, suggesting that the irradiation did not have a significant impact, if any, on the performance of Cu-OFE.

Cluster in breakdowns with respect to pulses occurred multiple times. To try to reduce the damage to the electrodes when clusters occurred, the target field was reduced to the previous stable point and pulsed before continuing to be increased. It can be seen from the conditioning chart shown in Figure 3 that the irradiated pair had a larger number of breakdowns at the start, which decreased during conditioning to a similar level as that of the non-irradiated pair, suggesting that it conditioned away any breakdown-inducing features caused by the irradiation.

Figure 4 displays the breakdown locations on the irradiated electrode where clusters occurred in different areas of the beam halo and not at all of the beam center area, suggesting that blisters are not the only cause of or may not have any effect on breakdowns. It is possible that the clusters on the halo are a result of the irradiation process as the non-irradiated electrode had no cluster of this sort, with breakdowns being dispersed over the surface.

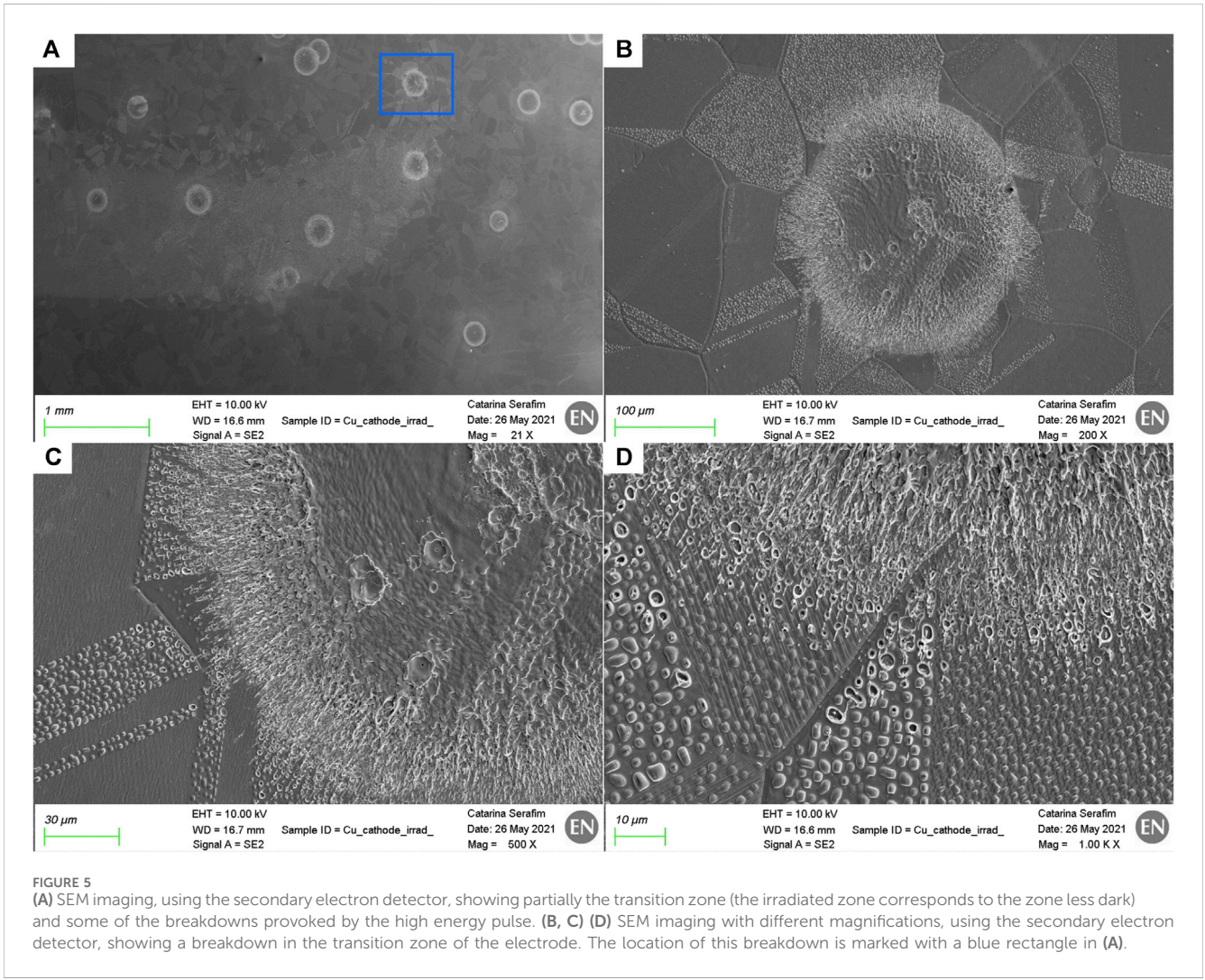

**FIGURE 5**
**(A)** SEM imaging, using the secondary electron detector, showing partially the transition zone (the irradiated zone corresponds to the zone less dark) and some of the breakdowns provoked by the high energy pulse. **(B, C) (D)** SEM imaging with different magnifications, using the secondary electron detector, showing a breakdown in the transition zone of the electrode. The location of this breakdown is marked with a blue rectangle in **(A)**.

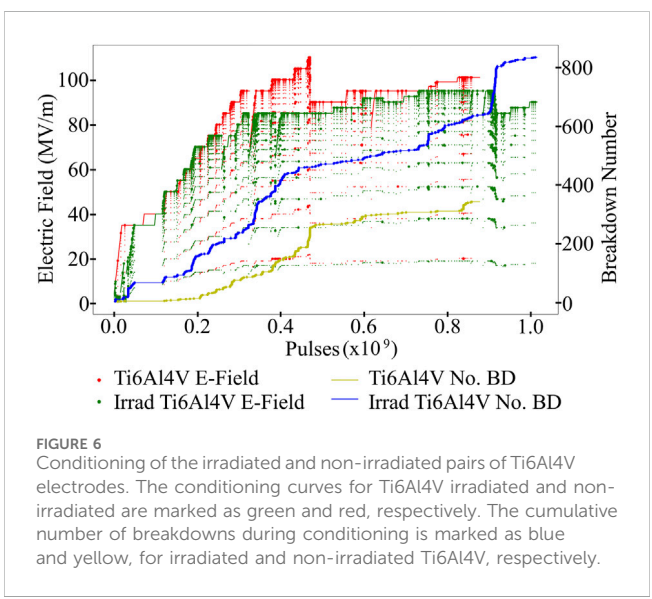

**FIGURE 6**
Conditioning of the irradiated and non-irradiated pairs of Ti6Al4V electrodes. The conditioning curves for Ti6Al4V irradiated and non-irradiated are marked as green and red, respectively. The cumulative number of breakdowns during conditioning is marked as blue and yellow, for irradiated and non-irradiated Ti6Al4V, respectively.

Observations from the SEM equipment confirmed that the low-energy beam irradiation caused the appearance of blisters on the Cu-OFE electrode's surface. These blisters are concentrated on the impact zone. It is visible that the blisters in the areas with the maximum deposition of beam have a higher density in quantity compared with zones where the beam was less intense.

Figure 5A–D show the results of Cu-OFE after the electrode had been irradiated with the $H^-$ beam and then subjected to high electric field pulsing in the LES. We can observe from Figure 5A that the breakdowns are well-distributed on the surface and do not seem to be correlated with the irradiated zone (zone less dark due to the blistering effect). Figures 5B–D show a breakdown in the transition zone, designated as the area where we start to see a decrease in the density of blistering. In this zone, it is visible that only some grains are affected by blisters. The fact that the phenomenon of blistering is more evident in just some of the grains may be correlated with the grain orientation of the material. Figures 5B, C also suggest that the

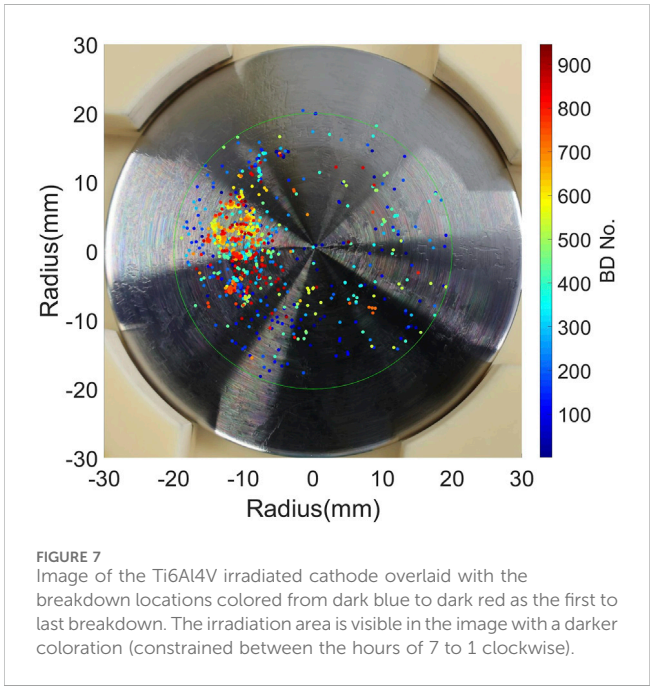

FIGURE 7
Image of the Ti6Al4V irradiated cathode overlaid with the breakdown locations colored from dark blue to dark red as the first to last breakdown. The irradiation area is visible in the image with a darker coloration (constrained between the hours of 7 to 1 clockwise).

grain orientation may play a role not only in the appearance of blisters but also in their emerging shapes. Additional studies will be performed to understand this correlation.

The breakdowns have a circular shape between approximately 100 μm and 200 μm, with splashes of melted material on their surroundings. The heat generated from the plasma formed during a breakdown has been shown to be significant enough to melt some of the blisters close to it, as shown in Figure 5D.

## 3.2 Titanium alloy (Ti6Al4V)

Concerning the LES experiment after irradiation, the Ti6Al4V electrodes have reached a stable field of 90 MV/m. Furthermore, the Ti6Al4V electrodes have shown a slight decrease in the breakdown performance due to irradiation—the pair of Ti6Al4V non-irradiated electrodes that was tested in the LES reached the average value of 100 MV/m. The conditioning of each pair can be seen in Figure 6.

Both pairs of Ti6Al4V electrodes exhibited a significant decrease in the operating field due to a temporal cluster of breakdowns that required a reduction in the electric field to reduce the risk of additional clusters. As this was a feature during the conditioning of both pairs, it would suggest that this is a property of Ti6Al4V rather than a spurious event. This may mean that Ti6Al4V has a tendency to accrue surface damage if conditioned too far. Overall, Ti6Al4V performed very well and was able to re-gain most of the previously achieved electric field without any further issues and maintain a stable field. The irradiated Ti6Al4V had a larger breakdown rate (BDR) at the start of conditioning, which then reduced, suggesting a strong conditioning effect of the irradiation area that improved the performance. Figure 7 shows the breakdown locations for the irradiated Ti6Al4V electrode, where the higher concentration of breakdowns appears to be constrained in the irradiation area.

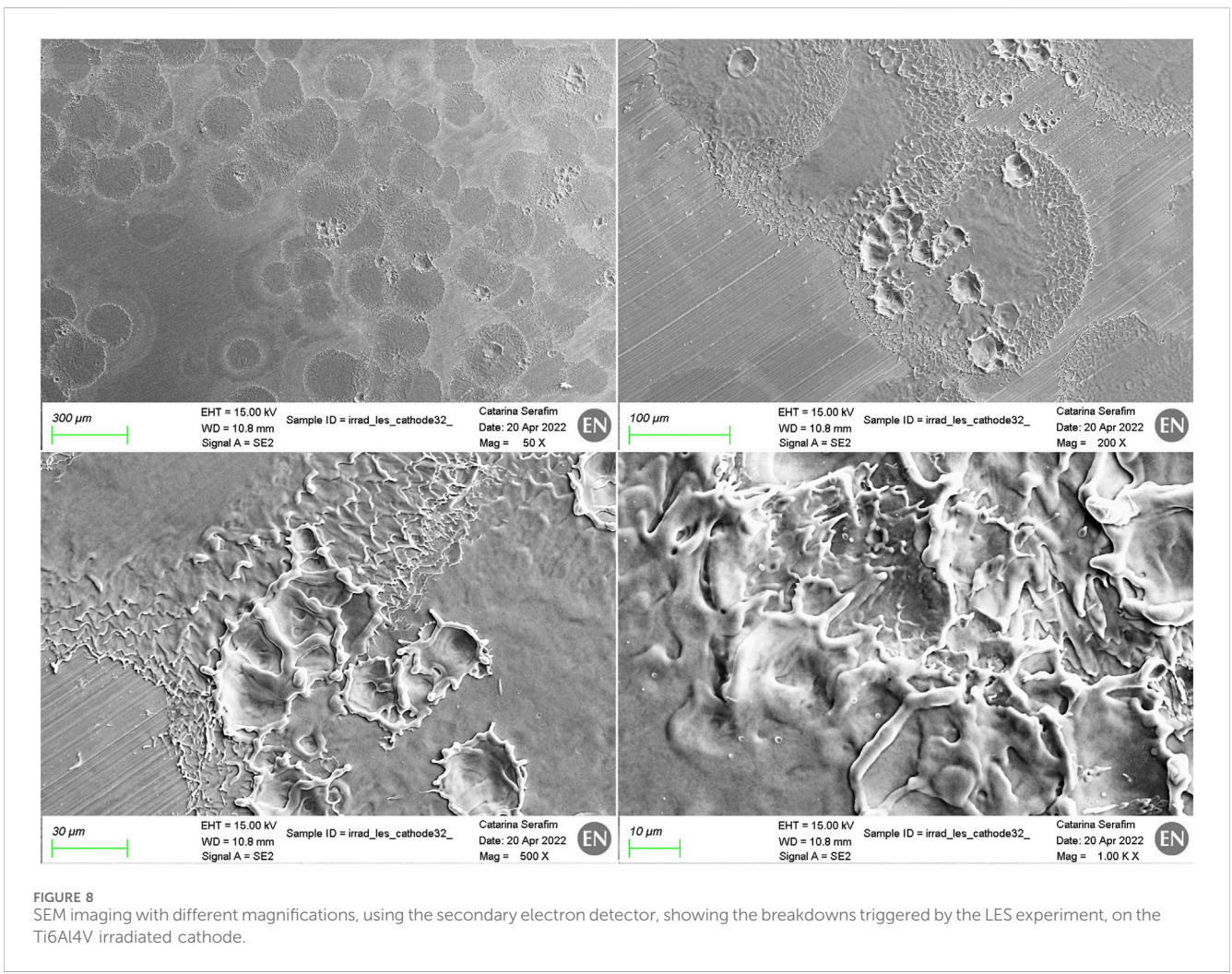

FIGURE 8
SEM imaging with different magnifications, using the secondary electron detector, showing the breakdowns triggered by the LES experiment, on the Ti6Al4V irradiated cathode.

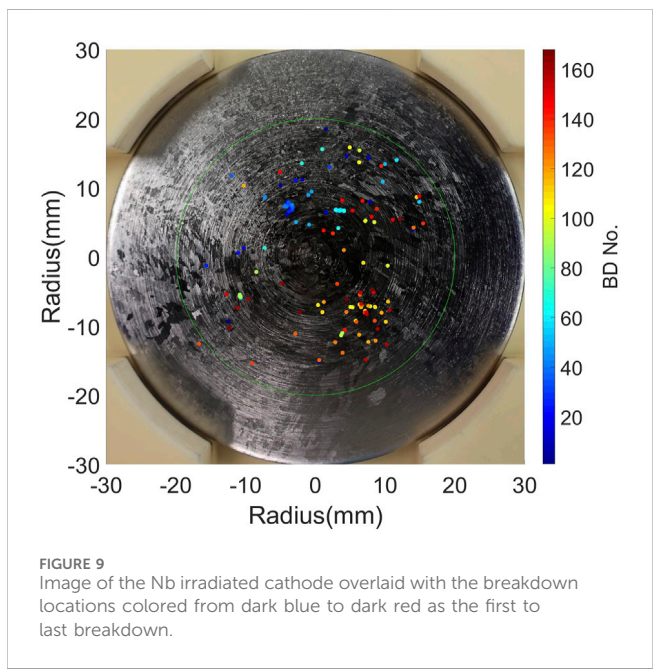

FIGURE 9
Image of the Nb irradiated cathode overlaid with the breakdown locations colored from dark blue to dark red as the first to last breakdown.

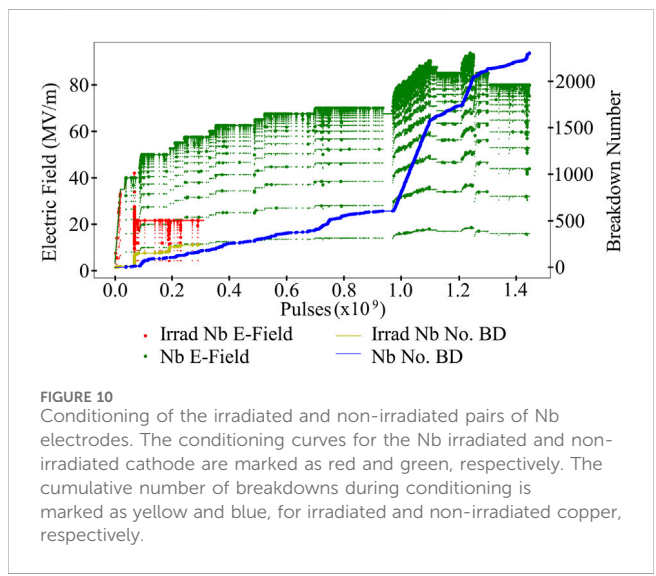

FIGURE 10
Conditioning of the irradiated and non-irradiated pairs of Nb electrodes. The conditioning curves for the Nb irradiated and non-irradiated cathode are marked as red and green, respectively. The cumulative number of breakdowns during conditioning is marked as yellow and blue, for irradiated and non-irradiated copper, respectively.

Figure 8 shows the breakdowns on the irradiated Ti6Al4V cathode on the surface. After irradiation, there was no evidence of modification of the surface structure due to irradiation nor appearance of blistering. Discoloration was seen, with a visible difference between the beam center, halo, and non-affected areas. The breakdown craters are similar in diameter to Cu, between 100 µm and 200 µm, due to the high field reached in the LES. No particular physical material defects were observed in these electrodes, which could explain why the irradiation area of the electrode presents a higher number of breakdowns.

### 3.3 Niobium (Nb)

Nb displayed similar results to those of Ti6Al4V with respect to physical defects and appearance after irradiation. The beam center area and halo for Nb can be seen in Figure 9. Figure 10 displays the conditioning plot for the Nb electrodes.

The irradiated pair had a large cluster of breakdowns upon reaching 42 MV/m during conditioning. This reduced the electric field, and the pair was unable to recover; they settled at a stable field of approximately 23 MV/m. The non-irradiated pairs did not experience any significant clusters and were able to achieve much higher fields, reaching a stable field of 80 MV/m. This considerable difference in the field values seems to have been caused by effects from the irradiation experiment. However, microscopy observations have shown no particular features in the surface or blistering effects that can be correlated with the decrease in performance. Investigations are ongoing to ascertain if this can be attributed to chemical surface modifications. Figure 11 shows some breakdowns on the Nb irradiated electrode. The breakdowns are smaller and less deep when compared with those of the Cu-OFE and Ti6Al4V electrodes, due to the smaller field achieved.

## 4 Conclusion

After irradiation with 45-keV $H^-$ particles, only Cu has shown blistering on the surface, as expected from hydrogen diffusivity. The affected areas where the blisters have appeared are indeed coincident with the footprint of the beam. During the high-voltage testing after irradiation, Cu and Ti6Al4V have shown the best performance in terms of reaching a stable field. Table 1 displays a summary of the materials tested, the maximum and stable field reached, and the BDR per pulse at the stable field given, where highlighted rows indicate the irradiation before testing. The stability at a specific field was determined based on the BDR, clusters in breakdowns with respect to pulses, occurrence of multiple breakdowns within a single pulse as determined by the cameras, and proximity to a previous limit causing a large cluster. It can be seen that all materials reached relatively high fields with only irradiated pairs being restricted, suggesting this as an effect of irradiation.

The good results from the non-irradiated and irradiated pairs of Ti6Al4V electrodes in the LES make us conclude that the irradiation did not induce any major decrease in performance. However, it may have led to some instabilities, causing a decrease in the obtainable field if run close to the limit. Taking only in consideration the experimental results from this study, it can be concluded that Ti6Al4V was the best material tested so far, with the highest number in both stable and maximum fields reached. However, taking into consideration that this material has lower thermal and electrical conductivities than copper, a careful design and a more complex assembly process would be required if used for an RFQ.

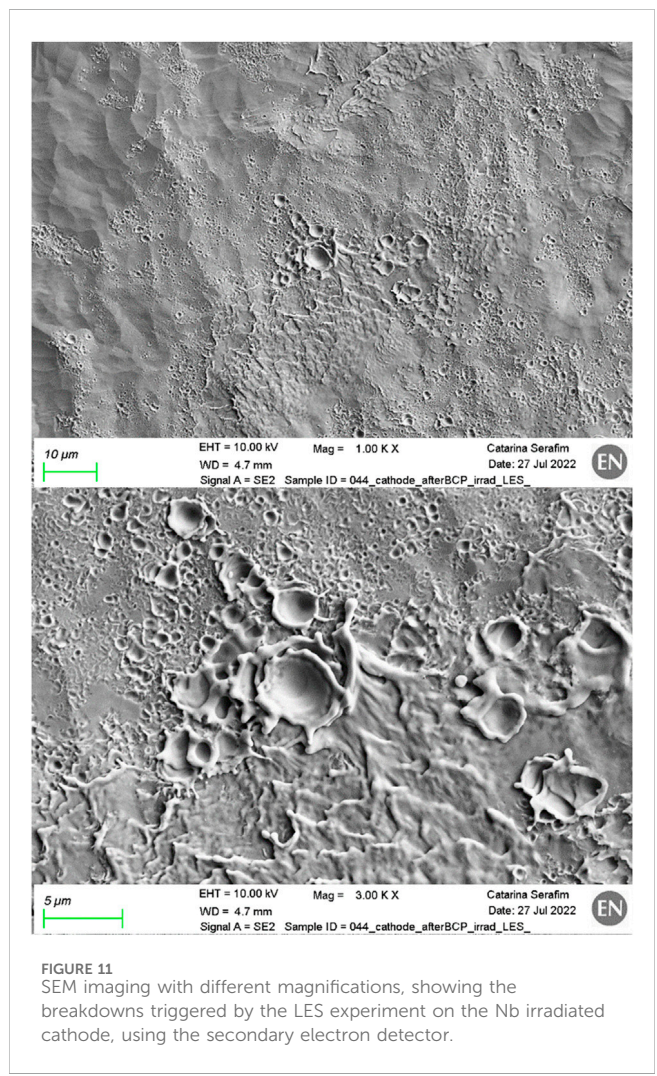

FIGURE 11
SEM imaging with different magnifications, showing the breakdowns triggered by the LES experiment on the Nb irradiated cathode, using the secondary electron detector.

TABLE 1 Summary of the stable, maximum field reached and the BDR at the stable field for each material. Rows highlighted indicate that the pair was irradiated before testing.

| Material | Stable field (MV/m) | Maximum field (MV/m) | Stable BDR |
|---|---|---|---|
| Ti6Al4V | 100 | 110 | $6.04 \times 10^{-7}$ |
|  | 90 | 95 | $1.58 \times 10^{-7}$ |
| Cu-OFE | 80 | 80 | $3.13 \times 10^{-7}$ |
|  | 80 | 80 | $7.5 \times 10^{-7}$ |
| Nb | 80 | 94 | $1.66 \times 10^{-6}$ |
|  | 21.7 | 42 | $4.09 \times 10^{-7}$ |

Copper gave the second-best results and appeared to reach the same field with an initial increase in the number of breakdowns in the irradiated electrode that appeared to condition away. This study suggests that the blistering phenomenon does not present a significant impact on the breakdown appearance or conditioning performance. Copper is currently used for the L4 RFQ and is well-established within the field; therefore, it would not require any manufacturing changes.

Nb had a significant reaction to the irradiation by causing a large and unpredictable cluster in breakdowns, from which it was not possible to recover from. The inferior performance of the material in terms of the reachable high field for the irradiated pair of electrodes, with the additional downside of being difficult to machine, leads us to exclude it from the list of possible materials for a future RFQ.

Finally, it should be underlined that in all the electrodes, some deposition of carbon on the surface was observed after irradiation. Further investigations are under way to better understand the source of the carbon and to avoid it in future experiments. If the carbon is inherent in the irradiation from the LINAC4 source, then it will also be expected in the operating conditions of the RFQ. These future studies will also try to assess whether any correlation exists between the presence of carbon in the electrodes surface and the breakdown phenomenon. Additionally, the discoveries observed regarding the different blisters' growth on the different grains of Cu-OFE are being further explored. As these electrodes have dimensions which are incompatible with those of the current EBSD equipment, dedicated samples are under study, with proper geometric dimensions, in order to be mounted in the dedicated chamber for irradiation and to perform EBSD analysis. This analysis will help us identify the orientation of each grain and further investigate how it affects blistering caused by $H^-$ ion implantation.

It should be noted that in the case of the electrodes, the entire irradiation was done before testing. On the other hand, in the RFQ, this occurs throughout running, and depending on whether irradiation occurs faster or slower than the rate of conditioning, it could influence the stability of the structure. This is most relevant for pairs of electrodes that reach similar electric fields when comparing irradiated and non-irradiated electrodes. Re-irradiating or constant irradiation may affect the performance or the achievable field. Tests are planned for conditioning irradiated electrodes and then re-irradiating and re-conditioning to see how this impacts the performance.

## Data availability statement

The original contributions presented in the study are included in the article/Supplementary Material; further inquiries can be directed to the corresponding author.

## Author contributions

CS: conceptualization, data curation, formal analysis, investigation, methodology, project administration, writing–original draft, and writing–review and editing. RP: Formal analysis, Investigation, Methodology, Writing–original draft, and Writing–review and editing. SC: project administration, supervision, validation, and writing–review and editing. Walter WW: project administration, supervision, validation, and writing–review and editing. FD: project administration, supervision, validation, and writing–review and editing. AF: investigation, resources, supervision, validation, and writing–review and editing. SS: resources, supervision, validation, and writing–review and editing. AG: project administration, resources, supervision, and writing–review and editing. AL: project administration, resources, and writing–review and editing. ES: project administration, resources, and writing–review and editing. SR: project administration, resources, and writing–review and editing. GB: project administration, resources, and writing–review and editing.

## Funding

The author(s) declare that no financial support was received for the research, authorship, and/or publication of this article.

## Acknowledgments

The authors thank all the members of the Work Package 10 of the L4 spare RFQ project. The authors also thank all the members of the MME team at CERN who have shared all the knowledge, open access to the microscopy facilities, and for the manufacturing of all the electrodes, and the VSC team for surface treatments of materials. Furthermore, the authors would like to thank the Hadron Sources and LINAC technicians who have provided access to the facilities and manipulated the procedure for irradiation.

## Conflict of interest

The authors declare that the research was conducted in the absence of any commercial or financial relationships that could be construed as a potential conflict of interest.

## Publisher's note

All claims expressed in this article are solely those of the authors and do not necessarily represent those of their affiliated organizations, or those of the publisher, the editors, and the reviewers. Any product that may be evaluated in this article, or claim that may be made by its manufacturer, is not guaranteed or endorsed by the publisher.